\documentclass[usenatbib]{mn2e}

\newcommand{\bbb}{\bigskip}
\newcommand{\be}{\begin{eqnarray}}

\newcommand{\bit}{\begin{itemize}}
\def\bkt#1{\left(#1\right)}
\def\bkts#1{\left[#1\right]}

\def\dd#1#2{{d{#1}\over d{#2}}}

\def\ee{\end{eqnarray}}
\def\eg{\textit{e.g.} }
\newcommand{\eit}{\end{itemize}}

\def\etal{\textit{et al.}}

\def\gtsima{$\; \buildrel > \over \sim \;$}

\def\ie{\textit{i.e.}}

\newcommand{\ii}{\textit}

\def\lab{\label}

\def\ltsima{$\; \buildrel < \over \sim \;$}

\def\mc#1{\mathcal{#1}}
\newcommand{\mmm}{\medskip}
\def\mpc{Mpc$^{-1}$ }

\newcommand{\no}{\noindent}
\newcommand{\nn}{\nonumber}

\def\pr{\prime}

\def\re#1{(\ref{#1})}

\def\simlt{\lower.5ex\hbox{\ltsima}}
\def\simgt{\lower.5ex\hbox{\gtsima}}
\def\sub#1{_{\mbox{\scriptsize{#1}}}}

\usepackage{amsmath}
\usepackage{hyperref}
\usepackage{amsfonts}
\usepackage{amssymb}
\usepackage{graphicx}
\usepackage{epsfig}
\DeclareMathAlphabet{\mathpzc}{OT1}{pzc}{m}{it}
\title[Imprints of dynamical dark energy on weak-lensing measurements]{Imprints of dynamical dark energy on weak-lensing measurements}
\author[Chongchitnan and King]{Sirichai Chongchitnan,$^1$	Lindsay King $^{2,3}$\\
	$^1$ Oxford Astrophysics, Denys Wilkinson Building, Keble Road, Oxford, OX1 3RH, United Kingdom,\\
	$^2$ Institute of Astronomy, Madingley Road, Cambridge, CB3 OHA, United Kingdom,\\
$^3$ Kavli Institute for Cosmology, Madingley Road, Cambridge, CB3 OHA, United Kingdom.}
\date{October 2009}

\pubyear{2010}

\begin{document}

\maketitle
\begin{abstract}
We show that simple models of scalar-field dark energy leave a generic enhancement in the weak-lensing power spectrum when compared to the $\Lambda$CDM prediction. In particular, we calculate the linear-scale enhancement in the convergence (or cosmic-shear) power spectrum for two best-fit models of scalar-field dark energy, namely, the Ratra-Peebles and SUGRA-type quintessence. Our calculations are based on linear perturbation theory, using gauge-invariant variables with carefully defined adiabatic initial conditions. We find that geometric effects enhance the lensing power spectrum on a broad range of scales, whilst the clustering of dark energy gives rise to additional power on large scales. The dark-energy power spectrum for these models are also explicitly obtained. On degree scales, the total enhancement may be as large as $30$-$40\%$ for sources at redshift $\sim1$. We argue that there are realistic prospects for detecting such an enhancement using the next generation of large telescopes.

\end{abstract}

\begin{keywords}
cosmology -- dark energy  -- gravitational lensing.
\end{keywords}
\section{Introduction}

Over the past years, a concordance picture of our universe has emerged from a number of cosmological probes.
In this paradigm, the dominant form of matter is cold and dark (CDM), but most of the mass-energy budget is in the form of dark
energy, manifest in the accelerated expansion of the universe [for some of the earliest evidence from  Type Ia supernovae, see for example  \cite{GoldPerl1998} and \cite{Riess1998Aj}]. Some of the key goals in cosmology from the analysis of upcoming high-precision experiments within the next couple of decades are to determine the nature of dark energy, and whether there is a tension with its interpretation as a cosmological constant.

A decisive way to distinguish whether dark energy is the cosmological constant or some other dynamical entity is to establish whether the dark-energy equation-of-state parameter $w\sub{DE}$ is constant or varies with redshift. The variation of $w\sub{DE}$ can be probed via a range of astronomical techniques, including measurements of light-curves of type Ia supernova (SNIa), cosmic microwave background (CMB) anisotropies, galaxy redshift surveys, galaxy cluster abundance and cosmological weak lensing  (see \cite{copeland,frieman+} for a summary of the various techniques).

Out of these techniques, weak lensing holds great promise in the understanding of dark energy, and has been identified as the most powerful individual technique, or most important component in a multi-technique study, provided systematic errors are well understood \citep{detf, peacock}. Its utility as a dark-energy probe comes from sensitivity to both the
expansion history of the universe and to the rate of growth of structures.

Observation of weak lensing is still in its infancy, with the first detection of cosmic shear made only nine years ago \citep{wittman}. At present, there are a number of upcoming ambitious experiments - both terrestrial [\eg LSST, ELT \citep{lsst, elt}] and space-based [\eg JDEM\footnote{http://jdem.gsfc.nasa.gov}, Euclid\footnote{http://sci.esa.int/euclid}], aimed at measuring cosmic shear to high accuracy. Given this rapid development, it is worth investigating if the upcoming weak-lensing experiments could discriminate simple models of dynamical dark energy from the cosmological constant. 

Many previous works on this problem concentrate solely on the geometrical effects of dynamical dark energy. Furthermore, many authors assume dynamical dark energy with $w\sub{DE}=$ constant, or some algebraic expression [numerous examples may be found the review by \cite{copeland}]. Whilst this is a useful approximation for studying cosmological dynamics at late-time, it is impossible to extrapolate these simple forms of $w\sub{DE}$ to early times, unless some \ii{ad hoc} mechanisms are invoked. As a result, one cannot consistently analyse the perturbations in dark energy, and many authors then neglect these perturbations altogether. Even those who do include dark-energy perturbations often gloss over the issue of initial conditions, and simply assume that the perturbations are insensitive to initial conditions without any justification.

In this work, we shall calculate the effects of dynamical dark energy on weak-lensing measurements without making any of the assumptions in the previous paragraph. We shall analyse simple models of dynamical dark energy using gauge-invariant perturbation theory, with well-defined initial conditions. As we shall see, these models leave some interesting imprints on the weak-lensing observables on all scales. Ultimately, we shall give a basic assessment of the prospects for distinguishing dynamical dark energy from the cosmological constant using weak-lensing measurements.

One particularly important aspect of this work is the inclusion of dark-energy perturbations throughout our analysis. There have been a number of recent works examining the effects of dark-energy perturbations on cosmological observables \citep{dent, sapone,hwang2}, including some interesting numerical simulations \citep{alimi, jennings}. Our approach is complementary to these works, and 
goes further in quantitatively establishing the effects of dynamical dark energy on weak-lensing power spectra. As a by-product, we obtain explicitly, for the first time, the dark-energy power spectrum for these models. Such a spectrum is useful as it quantifies the clustering of dark energy as a function of physical length scale.

Throughout this paper, we work in Planck units in which $c=\hbar=1$. 

\section{Dynamical Dark Energy}

One of the simplest and most widely studied models of dynamical dark energy is \ii{quintessence} - a scalar field evolving slowly along a potential $V(\phi)$  \citep{caldwell, ratrapeebles,wetterich,ferreira}. In this work, we shall focus on the weak-lensing signatures of two particular models of quintessence, namely, the Ratra-Peebles and the SUGRA models.

\subsection{The Ratra-Peebles potential}
The inverse power-law potential of the form
\begin{align} V(\phi)={M^{4+\alpha}\over\phi^\alpha}, \quad \alpha>0\end{align}
has been investigated by \cite{ratrapeebles} and \cite{wetterich}. In this model, the dark-energy density tracks the background energy density by mimicking the equation of state during radiation and matter epochs, and subsequently dominates only at late times. This tracking behaviour holds generally for a wide range of initial conditions.

The attractor properties depend on the slope $\alpha$ roughly as follows. When $\alpha$ is large, the potential becomes very flat for large values of $\phi$, hence effectively mimicking the cosmological constant at late times. However, the steep part of the potential for small values of $\phi$ tends to induce a prolonged matter-dominated epoch. Constraints from distance measurements tend to favour smaller values of $\alpha$, although the initial conditions then become increasingly difficult to adjust (note that the quintessence would require as much fine-tuning as the cosmological constant as $\alpha\rightarrow0$). A trade-off therefore has to be made in this respect. Given the slope, the self-coupling constant $M$ can then be set to match the value of the dark energy density today. Typically $M\ll m\sub{pl}$  and therefore there is inevitably a hierarchy problem associated with such a light scalar field. We do not concern ourselves with this issue here, but refer the reader to discussions elsewhere \citep{carroll,frieman}.

In this paper, we set the self-coupling constant $M=4.9$ eV and the slope $\alpha=0.5$, given by \cite{alimi}, who obtained these `best-fit' values from a likelihood analysis of the WMAP 5-year data \citep{wmap5} and the `union' SNeIa data \citep{kowalski} [see also earlier studies by \citep{klypin, solevi, dolag, maio, casarini} who investigated slightly different values of the parameters].

\subsection{The `SUGRA' potential}
This potential is of the Ratra-Peebles form augmented by an exponential term:
\begin{align} V(\phi)={M^{4+\alpha}\over\phi^\alpha}e^{4\pi G\phi^2}.\end{align}
\cite{brax} obtained this potential based on the motivation that a supersymmetric correction should be incorporated into the quintessence potential as $\phi$ can attain a large value exceeding a  Planck mass at late times. With this correction, the dynamical behaviour is now less sensitive to the value of $\alpha$, which is loosely constrained to be of of order $1$, and the scalar field need  not be as light as the Ratra-Peebles case. Again we shall use the values of the constants based on the analysis of Alimi \etal, namely $\alpha=1$ and $M=2.1\times10^3$ eV (see also the references cited above).

\section{Perturbations}

\subsection{Background}

We consider a background of a homogeneous, isotropic and spatially flat universe, which can be described by the Friedmann-Robertson-Walker metric
$$ds^2=-dt^2+a^2(t)dx^idx_i,$$
where $a(t)$ is the scale factor given in terms of cosmic time $t$. We assume that the universe contains only radiation, cold dark matter and dark energy, whilst baryons are neglected throughout. Baryons can alter the clustering of matter via baryonic cooling in dark matter halos \citep{white,rudd}. 
%%%%
This, however, only translates to small changes in the the weak-lensing spectra on scales very much smaller than those considered here. As for the geometric effect of quintessence on baryon acoustic oscillations, there are various degeneracies to overcome and preliminary results from $N$-body simulations do not appear optimistic \citep{jennings:2010fj}. For these reasons, we feel that it is reasonable to neglect baryons in this work.

With these assumptions, the Friedmann and acceleration equations relating the Hubble parameter, $H$, to the energy density, $\rho$, and pressure, $p$, are given by
\begin{align}&H^2= {\bar\kappa^2\over3}\rho={\bar\kappa^2\over3}\bkts{\rho_m+\rho_r+\rho_\phi},\lab{fried}\\
&\dot H =-{\bar\kappa^2\over2}\bkts{\rho_m+{4\over3}\rho_r+\rho_\phi+p_\phi},\lab{ray}\end{align}
where $\bar\kappa^2=8\pi G$.  As a convention, we use the subscripts $m$, $r$ and $\phi$ to denote quantities related to dark matter, radiation and dark energy respectively. The `total' quantities carry no subscript. An overdot indicates a derivative with respect to $t$ unless stated otherwise.

The quintessence field $\phi$ also satisfies the Klein-Gordon equation
\begin{align} 
\ddot{\phi}+3H\dot\phi+V^\pr(\phi)=0.\lab{kg} 
\end{align}

The various energy densities obey the conservation equations
\begin{align} 
\dot\rho_m+3H\rho_m =0,\lab{conser1}\\ 
\dot\rho_r+4H\rho_r =0,\lab{conser2}\\ 
\dot\rho_\phi+3H(\rho_\phi+p_\phi)=0.\lab{conser3}
\end{align}
The equation-of-state parameter $w$ is defined as the ratio $p/\rho$. For dark matter and radiation, we have $w_m=0$ and $w_r=1/3$ respectively. In the case of the cosmological constant, $w_\Lambda\equiv p_\Lambda/\rho_\lambda=-1$. For quintessence, we have
\begin{align} \rho_\phi = {1\over2}\dot\phi^2+V,\qquad p_\phi = {1\over2}\dot\phi^2-V, \qquad w_\phi\equiv {p_\phi\over \rho_\phi}.\lab{energy}\end{align}
The effective equation of state is given by
$$ w\equiv {p\over\rho}={\rho_r\over3\rho}+{\rho_\phi\over\rho}w_\phi.$$

In practice, it is more convenient to recast the background evolution equations \re{ray}--\re{conser3} in terms of the following dimensionless variables:
\begin{align} x\equiv {\bar\kappa\dot\phi\over \sqrt{6}H}, &\qquad y\equiv {\bar\kappa\over H}\sqrt{V\over3},\nn\\
\mathpzc{z}\equiv {\bar\kappa\over H}\sqrt{\rho_r\over3}, \qquad u\equiv {\bar\kappa\over H}&\sqrt{\rho_m\over3}, \qquad \lambda\equiv-{1 \over \bar\kappa V}\dd{V}{\phi}.\nn
\end{align}
This yields a set of differential equations:

\begin{align}
\dd{\ln H}{N}&= -{3\over2}\bkt{1+x^2-y^2+{\mathpzc{z}^2\over3}},\nn\\
\dd{x}{N}&= \sqrt{3\over2}\lambda y^2-x\bkt{3+\dd{\ln H}{N}},\nn\\
\dd{y}{N}&= -\sqrt{3\over2}\lambda xy-y\dd{\ln H}{N},\nn\\
\dd{\mathpzc{z}}{N}&= -\mathpzc{z}\bkt{2+\dd{\ln H}{N}},\nn\\
\dd{u}{N}&= -u\bkt{{3\over2}+\dd{\ln H}{N}},\nn
\end{align}
where  $N=\ln a$. In addition, $x,y,\mathpzc{z},u$ satisfy the Friedmann's constraint
\begin{align}x^2+y^2+\mathpzc{z}^2+u^2=1.\end{align}
The dimensionless system above is numerically more robust to integrate, and the dynamics in the phase-space $(x,y,\mathpzc{z},u)$ can be easily analysed \citep{copelandliddle}. 

In our numerical work, the initial values of the background variables $\{x,y,\mathpzc{z},u\}$ were chosen in such a way that the energy densities today are in broad agreement with the observed values: 
$$\Omega_{\phi,0}\simeq0.27,\quad \Omega_{r,0}\simeq 8.6\times 10^{-5}, \quad\Omega_{m,0}=1-\Omega_{\phi,0}-\Omega_{r,0},$$
[see \eg \cite{wmap5,kowalski,sdss2}]. For computational convenience, we shall work with the dimensionless density parameter $\Omega\equiv\rho/\rho_c$ (where the critical density $\rho_c\equiv3H_0^2/\bar\kappa^2$) instead of $\rho$. The subscript $0$ indicates the present-day value.

In terms of $\Omega$, Friedmann's equation \re{fried} becomes the constraint
$$ \Omega=\Omega_m+\Omega_r+\Omega_\phi,$$
with $\Omega=1$ at the present time. The density parameters redshift according to
\begin{align}
\Omega_m&=\Omega_{m,0}\,(1+z)^{3},\lab{redshift1}\\
\Omega_r&=\Omega_{r,0}\,(1+z)^{4},\lab{redshift2}\\
\Omega_\phi&=\Omega_{\phi,0}\,(1+z)^{3}\exp\Theta(z),\lab{redshift3}\\
\Theta(z)&=3\int_0^z{w_\phi(z^\pr)\over 1+z^\pr}dz^\pr\lab{redshift4}.\end{align}

%Finally, it is useful to define the \ii{adiabatic} dark-energy sound speed as
%$$ c_\phi^2\equiv {\dot{p}_\phi\over \dot{\rho}_\phi}=1+{2V^\pr\over3H\dot{\phi}}\,,$$
%and the effective adiabatic sound speed as
%$$
%c_s^2\equiv{\dot{p}\over \dot{\rho}}={4\rho_r+9\dot{\phi}^2c_\phi^2\over 9\rho_m +12\rho_r+9\dot{\phi}^2}\,,$$
% where our $c_s^2$ does not refer to the ratio $\delta p/\delta \rho$ sometimes defined elsewhere as the sound speed. The latter ratio equals unity in the case of quintessence with a canonical Lagrangian.

\subsection{Gauge-invariant variables} 

Working in Newtonian gauge and neglecting anisotropic stresses, vector and tensor perturbations, the Friedman-Robertson-Walker metric with scalar metric perturbation $\Phi$ takes the usual form

\begin{align} ds^2=-(1+2\Phi)dt^2+a^2(t)(1-2\Phi)dx^idx_i.\end{align}
The components of the energy-momentum tensor are
\begin{align} T^0_0=-\rho,\quad T^0_\mu=-{1\over k}(\rho+p)v_{,\mu},\qquad T^\nu_\mu=p\delta^\nu_\mu,\lab{components}\end{align}
where the total energy density $\rho$ decomposes into the background value ($\bar{\rho}_m+\bar{\rho}_r+\bar\rho_\phi$) and the perturbed part ($\delta\rho_m+\delta\rho_r+\delta\rho_\phi$) and $k$ is the wavenumber. The total pressure $p$ and velocity $v$ can be similarly decomposed. 
%% Derive integrated potential along line of sight, hence the power spectrum P_\phi

We define the density contrasts $\delta$ and the quintessence perturbation $\delta\phi$ by
\begin{align} \delta_m = {\delta\rho_m \over  \bar\rho_m },\quad\delta_r = {\delta\rho_r \over  \bar\rho_r },\quad \phi=\bar\phi+\delta\phi\,.\quad\end{align}
The total metric perturbation comprises contributions from all three components:
\begin{align} \Phi= \Phi_m+\Phi_r+\Phi_\phi,\end{align}
where each component relates to the scalar perturbations via Poisson's equation in Fourier space
\begin{align} \Phi_i={3\over2}\bkt{aH_0\over k}^2\bkt{\Omega_i\Delta_i},\lab{poisson}\end{align}
with $i=m, r,\phi$. Here, the gauge-invariant density variables $\Delta_i$ are given by 
\begin{align}\Delta_m &= \delta_m+3\bkt{aH\over k}v_m, \lab{gi1}\\
\Delta_r &= \delta_r+4\bkt{aH\over k}v_r,\lab{gi2} \\
\Delta_\phi&= { {\dot{\phi}\dot{\delta\phi}+[3H\dot{\phi}+V^\pr(\phi)]\delta\phi+{\dot\phi}^2\Phi}\over{{1\over2}{\dot{\phi}^2}+V(\phi)}}, \lab{gi3}\\
v_\phi&= {k\,\delta\phi\over a \dot{\phi}} \lab{gi4},\end{align}
\citep{kodama}. Note that in Newtonian gauge, it is $\Delta$, and not $\delta$, that appears in Poisson's equation. The variable $\Delta$ is velocity-dependent, but in practice one finds that the velocity effect is only noticeable on the largest scales.
%Kodama 1.37 Ch6

The gauge-invariant variables $\{\Delta_i,v_i\}$ evolve via a set of six coupled differential equations \citep{kodama}:

\begin{align}
\dd{\Delta_m}{z}=&{k\over H}v_m-{9\over2}{a^2H\over k}(1+w)(\bar{v}-v_m),\lab{d1}\\
\dd{\Delta_r}{z}+{\Delta_r\over1+z}=&{4k\over 3H}v_r-{6}{a^2H\over k}(1+w)(\bar{v}-v_r),\lab{d2}\\
\dd{\Delta_\phi}{z}+{3w_\phi\Delta_\phi\over 1+z}=&{\dot\phi^2\over\rho_\phi}\bkts{{k\over H}v_\phi-{9\over2}{a^2H\over k}(1+w)(\bar{v}-v_\phi)},\lab{d3}\\
\dd{v_m}{z}-{v_m\over 1+z}=&{3a^2H\over2k}\bar\Delta,\lab{v1}\\
\dd{v_r}{z}-{v_r\over 1+z}=&{3a^2H\over2k}\bar\Delta-{k\over 4H}\Delta_r,\lab{v2}\\
\dd{v_\phi}{z}-{v_\phi\over 1+z}=&{3a^2H\over2k}\bar\Delta-{k\over (1+w_\phi)H}\Delta_\phi,\lab{v3}
\end{align}

where
\begin{align}
\bar{v}&={1\over{\rho(1+w)}}\bkt{\dot\phi^2v_\phi+{4\over3}\rho_rv_r+\rho_mv_m},\\
\bar{\Delta}&={1\over\rho}\bkt{\rho_m\Delta_m+\rho_r\Delta_r+\rho_\phi\Delta_\phi}.
\end{align}

The power spectrum of component $i$ is defined as
 
\begin{align} {P}_{i}(k)\equiv{1\over (2\pi)^3}\langle|\Delta_i(k)|^2\rangle.\end{align}

Assuming that the primordial spectrum $P_m(k)$ scales with $k^{n_s}$, where $n_s$ is the scalar spectral index, it can be shown that   
\begin{align} P_m(k) =2\pi^2\delta_H^2{k^{n_s}\over H_0^{n_s+3}}\bkts{\Delta^+_m(z=0)\over\Delta^+_m(z=z_i)}^2 \lab{growing}\end{align}
where the term in brackets is the ratio of the growing-mode amplitudes of matter perturbations $\Delta_m^+$ at redshift $z=0$ and initial redshift $z_i$ [see, \eg \cite{dodelson}]. The amplitude $\delta_H$ specifies the normalization of the matter power spectrum at CMB scales. In particular, \cite{liddleparkinson} have computed a fitting function for $\delta_H$ using results from WMAP \citep{wmap1}. Under the assumption that there is no tensor mode, the fitting formula reads
\begin{align} \delta_H(k=0.05 \mbox{ \mpc})=1.927\times10^{-5}\exp[-1.24(1-n_s)].\lab{wmapfit}\end{align}
In our numerical work, we shall use the fitting formula above with $n_s=0.96$, and set the initial redshift at $z_i=10^9$.

\subsection{Initial conditions}
We expect the initial conditions for the perturbations in the various components to be set during inflation, and therefore it is reasonable to impose the adiabatic initial conditions on all perturbations. Adiabaticity specifies that at an initial time $t_0$, the following quantities vanish:

\begin{itemize}
\item[(A1)] $\mathcal{S}_{mr}$, the entropic perturbation between dark matter and radiation, 
\item[(A2)] $\mathcal{S}_{m\phi}$, the entropic perturbation between dark matter and quintessence,
\item[(A3)] $\mathcal{I}$, the intrinsic entropic perturbation of quintessence.
\end{itemize}

The entropic perturbations above are defined as
\begin{align}
\mathcal{S}_{mr}\equiv{\delta\rho_m \over\dot{\rho}_m}-{\delta\rho_r \over\dot{\rho}_r}=0,\\
\mathcal{S}_{m\phi}\equiv{\delta\rho_m \over\dot{\rho}_m}-{\delta\rho_\phi \over\dot{\rho}_\phi}=0,\\
\mathcal{I}\equiv{\delta\rho_\phi \over\dot{\rho}_\phi}-{\delta p_\phi \over\dot{p}_\phi}=0,\end{align}
where the perturbations in  dark-energy density and pressure satisfy
\begin{align} \delta\rho_\phi=\dot{\phi}\delta\dot{\phi}-\dot{\phi}^2\Phi+V^\pr\delta\phi,\\
\delta p_\phi=\dot{\phi}\delta\dot{\phi}-\dot{\phi}^2\Phi-V^\pr\delta\phi.\end{align}
Note that the adiabaticity between radiation and quintessence, namely
\begin{align}
\mathcal{S}_{r\phi}\equiv{\delta\rho_r \over\dot{\rho}_r}-{\delta\rho_\phi \over\dot{\rho}_\phi}=0,\lab{auto}\end{align}
is automatically satisfied.

In terms of the gauge-invariant variables $\{\Delta_i,v_i\}$, the conditions (A1)-(A3) yield respectively

\begin{align} 
\Delta_m-{3\over4}\Delta_r&=3\bkt{aH\over k}(v_m-v_r),\lab{adia1}\\
\Delta_m+3\bkt{aH\over k}^2\bar\Delta-{\rho_\phi\over\dot\phi^2}\Delta_\phi&=3\bkt{aH\over k}(v_m-v_\phi),\lab{adia2}\\
3\bkt{aH\over k}^2\bar\Delta&={\rho_\phi\over\dot\phi^2}\Delta_\phi.\lab{adia3}
\end{align} 
Equation \re{adia3} can be further combined with \re{adia2} to yield
\begin{align}
\Delta_m&=3\bkt{aH\over k}(v_m-v_\phi)\,. \lab{adia4}\end{align}
Inserting this into \re{adia1} gives
\begin{align} 
\Delta_r&=4\bkt{aH\over k}(v_r-v_\phi)\,.\lab{adia5}\end{align}
One can also easily verify that this equation can be obtained directly from \re{auto}. 

Once the adiabatic conditions (A1)-(A3) are imposed, there remains the freedom to choose three initial values from the set $\{\Delta_i,v_i\}$. Firstly, the initial value of $\Delta_m$ is chosen so that the amplitude of the matter power spectrum is normalised  on CMB scales according to equation \re{wmapfit}. Next, assuming that the velocity perturbations are initially small compared with the density perturbations, we take $\Delta_r=4\Delta_m/3$ (suggested by the usual adiabatic condition for matter and radiation). Lastly, we set $v_\phi=0$ initially, as the dark-energy velocity perturbations quickly redshift away and therefore do not have a significant effect the growing mode amplitude. With these choices, all the initial values of $\{\Delta_i,v_i\}$ can be determined by solving the linear system consisting of equations \re{adia3},\re{adia4} and \re{adia5}.

It is important to note that adiabaticity implies ${\Delta_\phi\neq0}$ initially (\eg from equation \ref{adia3}), and hence dark-energy perturbations cannot be consistently neglected.

%@^@^@^@^@^@^@^@^@@^@^@^@^@^@^@^@^

\subsection{Power spectra}\lab{powerful}

%%%%%%%%%%%%%%%%%%%%%%%%%%
%%%%%%%% Figure 1 %%%%%%%%
%%%%%%%%%%%%%%%%%%%%%%%%%%

\begin{figure}
\centering
\epsfig{file=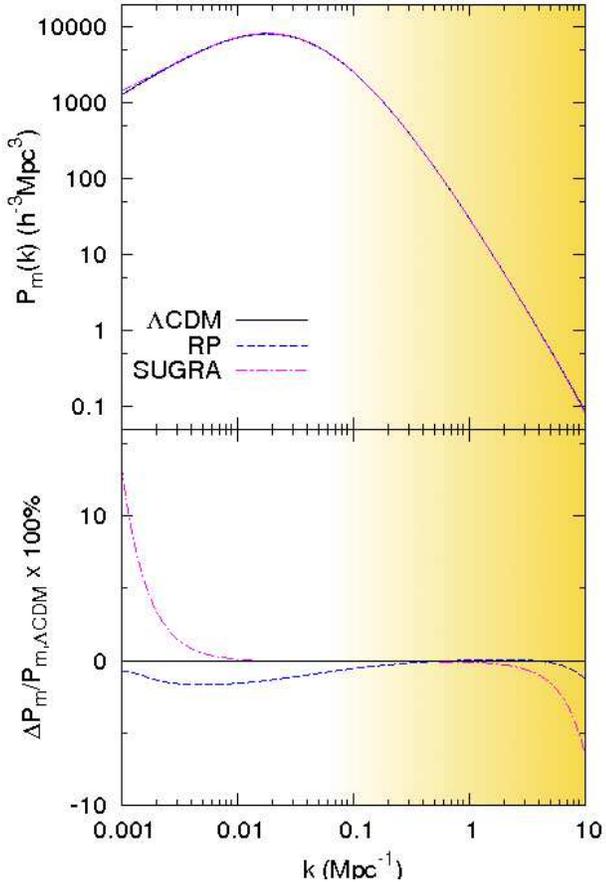, width= 12cm, angle = -90}
\caption{The linear matter power spectra $P_m(k)$ for the case that dark energy is the cosmological constant (solid/black line), the Ratra-Peebles quintessence (dashed/blue) or the SUGRA-type quintessence (dash-dot/magenta). The lower panel shows the fractional difference in $P_m(k)$  expressed as a percentage (equation \ref{fracdiff}). Non-linear effects manifest in the shaded region starting from $k\sim0.1\mbox{ Mpc}^{-1}$ (though there are other estimates as low as $0.03h\mbox{Mpc}^{-1}$ ). See text for further discussion. }
\label{figpm}
\end{figure} 

%%%%%%%%%%%%%%%%%%%%%%%%%%
%%%%%%%%%%%%%%%%%%%%%%%%%%
%%%%%%%%%%%%%%%%%%%%%%%%%%

Figure \ref{figpm} shows the linear matter power spectra $P_m(k)$ for the $\Lambda$CDM, Ratra-Peebles and SUGRA models. The lower panel shows the fractional difference in the matter power spectra between  the $\Lambda$CDM and the quintessence models, expressed as the ratio
\begin{align} {{P\sub{m,Quintessence}-P_{\mbox{\scriptsize{m}},\Lambda\mbox{\scriptsize{CDM}}}}\over P_{\mbox{\scriptsize{m}},\Lambda\mbox{\scriptsize{CDM}}} }\times100\%.
\lab{fracdiff}
\end{align}

The spectra in the upper panel are difficult to distinguish by eye, but the lower panel shows that there are some differences on large scales, where the SUGRA best-fit model yields an excess power of around $10\%$ compared to the $\Lambda$CDM model. On the other hand, the Ratra-Peebles best-fit model shows a very slight deviation in power from the $\Lambda$CDM case. In both cases, there is a slight suppression in the matter power spectrum on all scales smaller than the turnover scale, corresponding to the Hubble radius at the time of matter-radiation equality. Generally, the differences are inconspicuous, and thus it would be very difficult to distinguish between these models via galaxy redshift surveys and the abundance of massive galaxy clusters.

%% dark energy spectra! %%
%%%%%%%% Figure 2 %%%%%%%%
%%%%%%%%%%%%%%%%%%%%%%%%%%

\begin{figure}
\centering
\epsfig{file=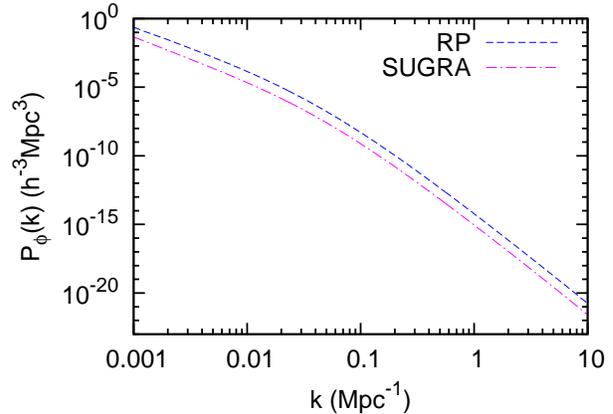, width= 6cm, angle = -90}
\caption{The linear dark-energy power spectra $P_\phi(k)$ for the Ratra-Peebles quintessence and the SUGRA-type quintessence. The spectra are roughly broken into two exponentially decaying regimes, changing around the turnover scale.}
\label{figde}
\end{figure} 

%%%%%%%%%%%%%%%%%%%%%%%%%%
%%%%%%%%%%%%%%%%%%%%%%%%%%
%%%%%%%%%%%%%%%%%%%%%%%%%%

We also obtained the power spectrum of dark energy $P_\phi(k)$, as shown in figure \ref{figde}. The amplitudes of these spectra are about 3 orders of magnitude below that of the matter power spectrum, thus showing that the clustering of dark energy is relatively weak. The spectra are roughly broken into two exponentially decaying regimes, changing around the turnover scale (the decay rate is greater for scales smaller than the turnover scale). The clustering power becomes negligible at sufficiently small scales as one might expect. Despite the small amplitude of $P_\phi$, the cross-correlation between the matter and the dark energy spectra yield a small but non-negligible contribution to the weak-lensing signal, as we shall see in the next section.
 
%in agreement with the findings of \cite{avelino, sapone}.

On smaller scales, one expects a significant modification of the various power spectra due to the non-linear evolution of the perturbations. $N$-body simulations suggest that non-linear effects are important for scales $k\gtrsim\mc{O}(0.1 \mbox{ Mpc}^{-1})$ , where corrections are of order $\sim10\%$ \citep{alimi, ma, jennings}. (It is worth noting an estimate in \citet{jennings:2010qy} which shows non-linearities manifesting on a much larger scale -- at around $k\sim0.03 h\mbox{Mpc}^{-1}$.) As far as standard non-linear fitting formulae for matter power spectrum are concerned [\eg \cite{peacockdodds,smith}], they are unreliable in the case when dark energy is dynamical, and they certainly do not shed any light on the non-linear clustering of dark energy. Thus, at present non-linear effects are usually calibrated against $N$-body simulations. This approach is clearly inefficient as they can only test a specific combination of parameters for a specific potential at a time. Some authors have turned to semi-analytic approaches to reproduce non-linear effects \citep{benabed,casarini} but these have not yet been tested sufficiently. From our point of view, combining simulations with field-theoretic calculations such as those developed by \cite{crocce} or \cite{bernardeau} may be the most sensible approach in dealing with the strongly non-linear regime. These analytic techniques are still being developed and, as far as we are aware, they are not yet applicable to models with dynamical dark energy. It is possible to calculate perturbations in quintessence cosmology analytically in a weakly non-linear regime using higher-order perturbation theory \citep{hwang, malik}, but these methods are very tedious and the results would be valid only on a narrow range of non-linear scales. We shall leave the issue of non-linearity for future investigation.

\section{Weak-lensing signals}

\subsection{The convergence spectrum}
The weak gravitational lensing of a light source may be quantified by two fields: the convergence $\kappa$, and the gravitational shear $\gamma$ [see \cite{bartelmann} for a review]. The convergence magnifies sources isotropically, and is therefore not directly observable unless the size of the source is known \ii{a priori}. In the weak lensing limit, the power spectrum of the convergence can be shown to be equivalent to the power spectrum of the cosmic shear, which is observable as a correlated distortion in the (complex) ellipticities of distant galaxies (ignoring any intrinsic alignment). In this work, we shall focus on the convergence power spectrum, although it can be equivalently replaced by that of the shear.

The convergence associated with a light source at comoving distance $r$ in the direction $\vec\theta$, resulting from the cumulative effect of density fluctuations at  comoving distance $r'$ is given by
%with 1/c^2 in front
\begin{align} \kappa(\vec\theta,r)=\int_0^rdr^\pr \bkt{{r-r^\pr}\over r}r^\pr\nabla^2\Phi(r^\pr\vec\theta,r^\pr).\end{align}
By integrating over all sources up to the Hubble radius $r_H$, the effective convergence in the direction $\vec\theta$ is

%with 1/c^2 in front
\begin{align} \kappa(\vec\theta)=\int_0^{r_H}dr \,W(r) \,r\,\nabla^2\Phi(r\vec\theta,r).\lab{keff}\end{align}
Here, the weight function 
\begin{align} W(r)=\int_r^{r_H}dr^\pr G(r^\pr)\bkt{r^\pr-r \over r^\pr},\end{align}
contains information on the distribution of sources along the line of sight. The function $G(r)$ is related to the source redshift distribution $p_z(z)$ by $G(r)dr=p_z(z)dz$.  In this work, we shall only consider sources at a single redshift, so that the weight function reduces to 
$$W(r)= \begin{cases} 1-{r/r^*} & \mbox{if } 0\leq r<r^*, \\ 0 & \text{otherwise,}\end{cases}$$
where $r^*$ is the comoving distance to the source.

The convergence power spectrum can be obtained by following essentially the same the derivation as in \cite{bartelmann}, but taking care of the dark-energy contribution to the total metric perturbation. The procedure involves using \re{redshift1}-\re{redshift4},\re{poisson}, \re{keff} and applying Limber's equation in Fourier space. We also neglect the contribution from radiation since $\Omega_{r,0}$ is small compared with $\Omega_{m,0}$ and $\Omega_{\phi,0}$. The convergence power spectrum, $P_\kappa$, expressed as a function of the multipole number, $\ell$, is given by
%factor 1/c^4 in front
\begin{multline} {P}_\kappa(\ell)\simeq{9H_0^4\over 4}\int_{0^+}^{r_H}{W^2(r)\over a^2}\bigg[\Omega_{m,0}^2{P}_m\bkt{{\ell\over r}, r}+\\ 
2\Omega_{m,0}\Omega_{\phi,0}\exp(\Theta)\,{P}_{m\phi}\bkt{{\ell\over r}, r}+\Omega_{\phi,0}^2\exp(2\Theta)\,{P}_\phi\bkt{{\ell\over r}, r}\bigg]dr,\lab{pk}\end{multline}
with $\Theta$ given by equation \re{redshift4}. By setting ${P}_\phi={P}_{m\phi}=0$, we recover the usual convergence power spectrum of Bartelmann \& Schneider.

The power spectra $P_i(k,r)$ in the integrand of \re{pk} are evaluated at scale $k$ for which $k=\ell/r$ and at redshift $z$ calculated from the  distance-redshift relation
\begin{align} r(z)= \int_0^z {dz^\pr\over H(z^{\pr})}. \lab{rofz}\end{align}
The range of integration extends to the comoving horizon $r_H=\lim_{z\rightarrow\infty}r(z)$.

%\section{Alternative version}
%Starting with \re{keff}, we use the metric perturbation derived in \cite{dent} based on earlier works by \cite{bartolo}
%\be
%\label{phiz}
%\frac{d^2\Phi}{dz^2} &=& - \frac{1}{1+z}\frac{d\Phi}{dz}\left(\frac{3}{2}w 
%- 3c_{a}^2 - \frac{3}{2}\right)\nn\\
%&&-\Phi\left(\frac{c_{a}^2 k^2}{H(z)^2} - 3\frac{(c_{a}^2 -w)}{(1+z)^2}\right) 
%+\frac{3}{2}\frac{\Omega_{\phi}}{\(1+z\)^2}\nn\\
%&&\times\[-c_{a,\phi}^2 S+\(1-c_{a,{\phi}}^2\)\Gamma\]\\
%\frac{dS}{dz}&=&-\frac{1}{1+z}\bigg[\(3w_{\phi}-\frac{3\Omega_{m}c_{a,{\phi}}^2}{1+w}\)S\nn\\
%   &&+\frac{3\Omega_{m}\(1-c_{a,\phi}^2\)}{1+w}\Gamma  +\frac{k^2 \(1+z\)^2}{H^2}\(\frac{1}{3}S+\frac{1}{3}\Gamma\)\nn\\
%   &&+\frac{k^4 \(1+z\)^4}{H^4}\(\frac{2}{9}\frac{\(1+w_{\phi}\)}{\(1+w\)}\Phi\)\bigg]\\
%\frac{d\Gamma}{dz}&=&-\frac{1}{1+z}\bigg[-\frac{3}{2}\(1+w\)S+3\(w_{\phi}-\frac{1+w}{2}\)\Gamma\nonumber\\
%&&- \frac{k^2 \(1+z\)^2}{H^2}\( \(1+w_{\phi}\){\cal R}+\frac{1}{3}S+\frac{1}{3}\Gamma\)\nn\\
%&&-\frac{ k^4 \(1+z\)^4}{H^4}\(\frac{2}{9}\frac{\(1+w_{\rm {DDE}}\)}{\(1+w\)}\Phi\)\bigg] \,.
%\end{align}

\subsection{Imprints of dynamical dark energy on the convergence}

%%%%%%%%%%%%%%%%%%%%%%%%%%
%%%%%%%% Figure 3 %%%%%%%%
%%%%%%%%%%%%%%%%%%%%%%%%%%

\begin{figure}
\centering
\epsfig{file=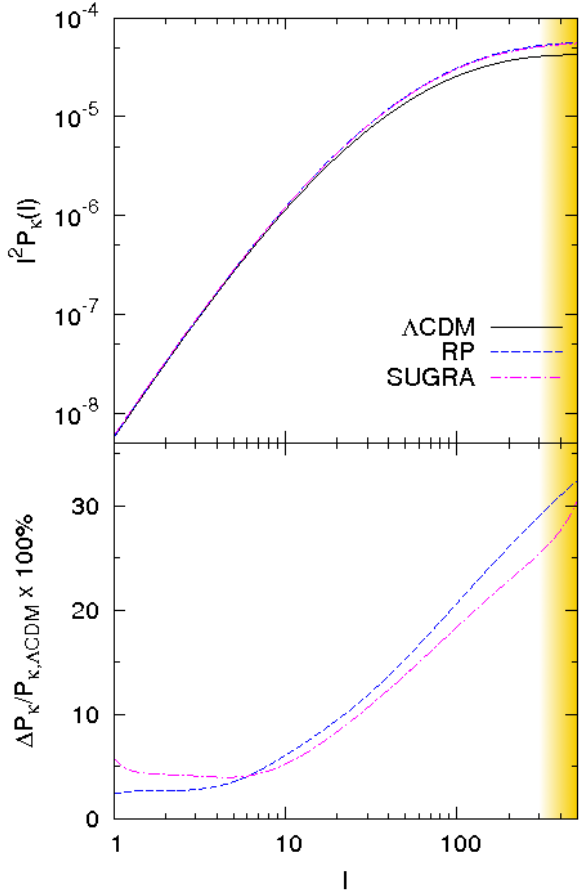, width= 12cm, angle = -90}
\caption{The convergence power per logarithmic $\ell$ interval for the case that dark energy is the cosmological constant (solid/black line), Ratra-Peebles quintessence (dashed/blue) or the SUGRA-type quintessence (dash-dot/magenta) assuming sources at $z=1$. The lower panel shows the fractional enhancement in $P_\kappa$ due to dynamical dark energy, expressed as a percentage. In the shaded region (beyond $\ell$ of a few hundred), non-linear effects become important. Further discussion is given in the text.}
\label{figkappa}
\end{figure} 

%%%%%%%%%%%%%%%%%%%%%%%%%%
%%%%%%%%%%%%%%%%%%%%%%%%%%
%%%%%%%%%%%%%%%%%%%%%%%%%%

Figure \ref{figkappa} shows the convergence power spectra for the cases in which dark energy is the cosmological constant, Ratra-Peebles or SUGRA-type quintessence, assuming sources at $z=1$. The upper panel shows $\ell^2P_\kappa(\ell)$, the total convergence power per unit logarithmic-$\ell$ interval. The lower panel shows the percentage difference in the convergence power for the quintessence models compared with the $\Lambda$CDM model. The convergence power spectra for the quintessence models are generally enhanced (by around $20\%$ at $\ell\simeq100$ or $\sim2^\circ$) even though there are relatively little differences in the matter power spectra. The enhancement occurs on all scales and in fact \ii{grows} with $\ell$.

\begin{figure}
\epsfig{file=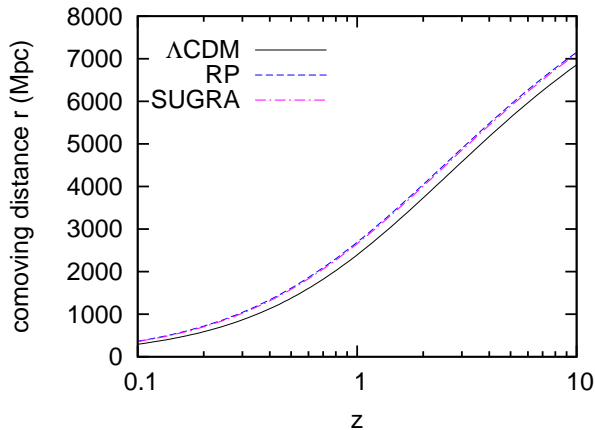, width= 6cm, angle = -90}
\caption{Distance-redshift relations given cosmology in which dark energy is the cosmological constant, Ratra-Peebles or SUGRA-type quintessence. Objects at a fixed redshift are further away in the quintessence cases, partially giving rise to the enhancement seen in figure \ref{figkappa}.}
\label{figdistance}
\end{figure}

\begin{figure}
\epsfig{file=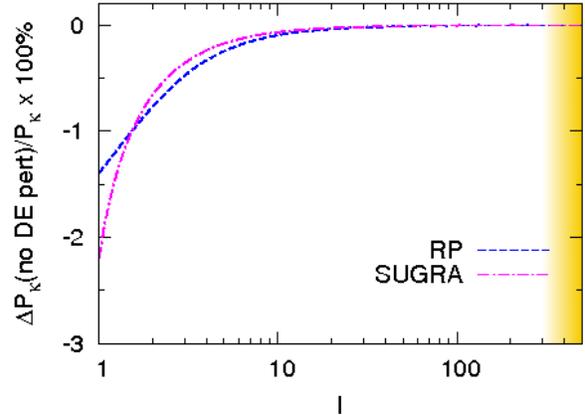, width= 6cm, angle = -90}
\caption{Percentage of error in $P_\kappa(\ell)$ if dark-energy perturbations were not included in the Newtonian potential $\Phi$ (\ie, ignoring the last two terms in equation \ref{pk}). Non-linear region is shaded as in figure \ref{figkappa}.}
\label{figignore}
\end{figure}

There are two main effects which contribute to the enhancement of the convergence power. Firstly, dynamical dark energy modifies the distance-redshift relation in such a way that a source at a fixed redshift lies at a greater comoving distance compared with the $\Lambda$CDM case  (see figure \ref{figdistance}). More intervening energy density therefore gives rise to a higher weak-lensing signal. Secondly, dark-energy perturbations contribute to an additional power via the last two terms in equation \ref{pk}. The cross- and auto-correlation of dark-energy perturbations together contribute an enhancement of a few percent on large scales $(\ell\lesssim10)$, as shown in figure \ref{figignore}. The $y$-axis is the ratio
\begin{align} {{P_{\kappa, \mbox{\scriptsize{no DE pert}}}-P_\kappa}\over P_\kappa}\times100\%,
\lab{fracignore}
\end{align}
where $P_{\kappa, \mbox{\scriptsize{no DE pert}}}$ is the convergence power spectrum calculated without the last two terms in equation \ref{pk}.

%%%%%%%%%%%%%%%%%%%%%%%%%%
%%%%%%%%%%%%%%%%%%%%%%%%%%

%%%%%%%%%%%%%%%%%%%%%%%%%%
%%%%%%%%%%%%%%%%%%%%%%%%%%
%%%%%%%%%%%%%%%%%%%%%%%%%%
%%%%%%%%%%%%%%%%%%%%%%%%%%

%%%%%%%%%%%%%%%%%%%%%%%%%%
%%%%%%%%%%%%%%%%%%%%%%%%%%
%%%%%%%%%%%%%%%%%%%%%%%%%%

\begin{figure}
\epsfig{file=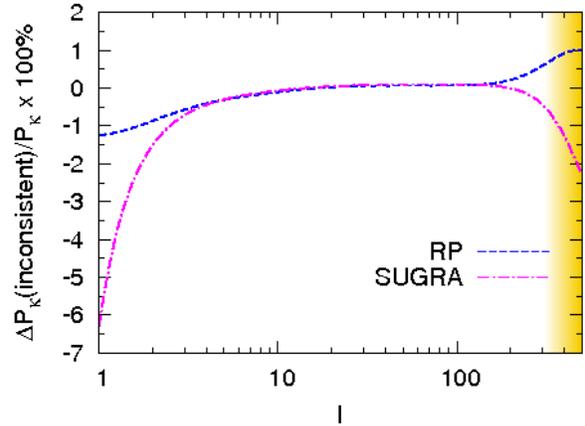, width= 6cm, angle = -90}
\caption{Percentage error in $P_\kappa(\ell)$ if dynamical dark energy were treated as though it does not cluster. The inconsistency is manifest in the divergence on small scales. Non-linear region is shaded as in figure \ref{figkappa}.}
\label{figwrong}
\end{figure}

One might ask what would happen to the convergence power spectrum if dark-energy perturbations were neglected \ii{altogether}, \ie, if we were to set $\Delta_\phi=v_\phi\equiv0$. This, however,  is an ill-posed problem as all dynamical models of dark energy must cluster, unless $w\equiv-1$ at all times. Neglecting this clustering gives rise to inconsistencies in the evolution equations, and the results will be sensitive to the gauge choice  \citep{hwang2}. In figure \ref{figwrong}, we plot the percentage `error' 
\begin{align} {{P_{\kappa\mbox{\scriptsize{, inconsistent}}}-P_\kappa}\over P_{\kappa} }\times100\%.
\end{align}
The figure shows that the error from ignoring dark energy clustering are of order $\lesssim10\%$, and can even grow at small scales. Nevertheless, this result is gauge-dependent, and therefore one must refrain from drawing any physical interpretation from this figure. It only serves as a cautionary example against ignoring the clustering of dark energy in the calculation of perturbations.

%A source at $z=1$ corresponds to a comoving distance of about $2$ Gpc. 

%%%%%%%%%%%%%%%%%%%%%%%%%%
%%%%%%%%%%%%%%%%%%%%%%%%%%

%%%%%%%%%%%%%%%%%%%%%%%%%%
%%%%%%%%%%%%%%%%%%%%%%%%%%
%%%%%%%%%%%%%%%%%%%%%%%%%%

On smaller scales, we expect an additional enhancement due to non-linear effects, coming solely from the change in geometry since dark energy does not cluster on such small scales. In the $\Lambda$CDM model, it is estimated that linear lensing spectra become unreliable with error $\sim10\%$ on scales beyond $\ell\sim\mbox{few}\times100$ \citep{jain,bartelmann}, and similarly for models with constant $w\sub{DE}$  \citep{huterer,refregier}. This is the reason that we only display our weak-lensing spectra up to $\ell=500$. Non-linear lensing spectra would certainly be a powerful method of distinguishing between models of dark energy, but as far quintessence are concerned, the non-linear lensing spectra has yet to be understood quantitatively. It would be interesting, though challenging, to investigate this problem using one of the methods described in section \ref{powerful}. Our present work nevertheless offers a first step towards understanding the signature of scalar-field dark energy on the weak-lensing spectra.

\section{Observational prospects}

We now consider if an enhanced weak-lensing signal due to dynamical dark energy is potentially observable by future weak-lensing surveys.

The error on the convergence power spectrum arising from cosmic variance and from the intrinsic dispersion in the ellipticities of the galaxies (from which the cosmic shear signal is measured) was derived by \cite{kaiser}:

\begin{align} \Delta P_\kappa(\ell)=\sqrt{2\over (2\ell+1) f\sub{sky}}\bkt{P_\kappa(\ell)+{\langle{\gamma\sub{int}^2\rangle}\over \bar{n}}}\lab{intrin}.
\end{align}
Here $f_{sky}$ is the fraction of the sky covered by a survey of area $\cal A$ (i.e. $\pi {\cal A}/129,600$deg$^{2}$), the mean-square intrinsic ellipticity is $\langle{\gamma\sub{int}^2\rangle}\approx0.16$ and the number of galaxies per steradian for which an ellipticity can be measured is $\bar{n}$.

%%%%Look
We consider two specific configurations: (A) ${\cal A}=20,000$\,deg$^{2}$ and a galaxy number density of 250\,arcmin$^{-2}$ and (B) ${\cal A}=30,000$\,deg$^{2}$ and a galaxy number density of 500\,arcmin$^{-2}$. Both configurations require deep imaging using one (A) or more (B) large ground-based telescopes such as the planned European Extremely Large Telescope (E-ELT) \citep{elt}, or using future space-based telescopes. 

\begin{figure}
\epsfig{file=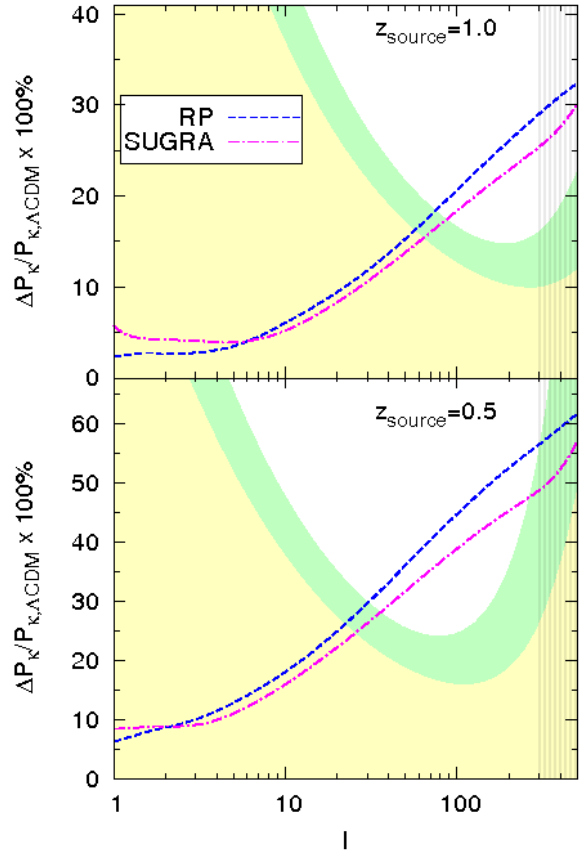, width= 12cm, angle = -90}
\caption{Percentage enhancement in the convergence power for the Ratra-Peebles and SUGRA quintessence, compared with that in the $\Lambda$CDM case. The error contours are calculated using equation \ref{intrin}, with (A) ${\cal A}=20,000 \mbox{ deg}^2$, galaxy number density = 250\,arcmin$^{-2}$ (outer contour) and (B) ${\cal A}=30,000 \mbox{ deg}^2$, galaxy number density = 500\,arcmin$^{-2}$ (inner contour). The sources are at $z=1$ (top panel) and at $0.5$ (bottom panel). Portions of the curves which lie above the shaded contours represent a potentially detectable enhancement. Note that non-linear effects become increasingly important for $\ell\gtrsim\mbox{few}\times100$ (grey, striped region), as discussed in the text.}
\label{figerrors}
\end{figure}

Figure \ref{figerrors} shows the errors \re{intrin} associated with these configurations [the outer contour corresponding to configuration (A)], for sources at redshift $1$ (top panel) and $0.5$ (bottom panel). For sources at $z=1$, the dark energy enhancement is obscured by the error in measurement below $\ell\simeq100$, and thus one must look for such a signal at smaller scales. For sources at $z=0.5$, the prospects are better, with dark-energy enhancement of around $40\%$ at $\ell\simeq100$, where the errors are minimum.

%%%%%%%%%%%%%%%%%%%%%%%%%%
%%%%%%%%%%%%%%%%%%%%%%%%%%

%%%%%%%%%%%%%%%%%%%%%%%%%%
%%%%%%%%%%%%%%%%%%%%%%%%%%
%%%%%%%%%%%%%%%%%%%%%%%%%%

\begin{figure}
\centering
\epsfig{file=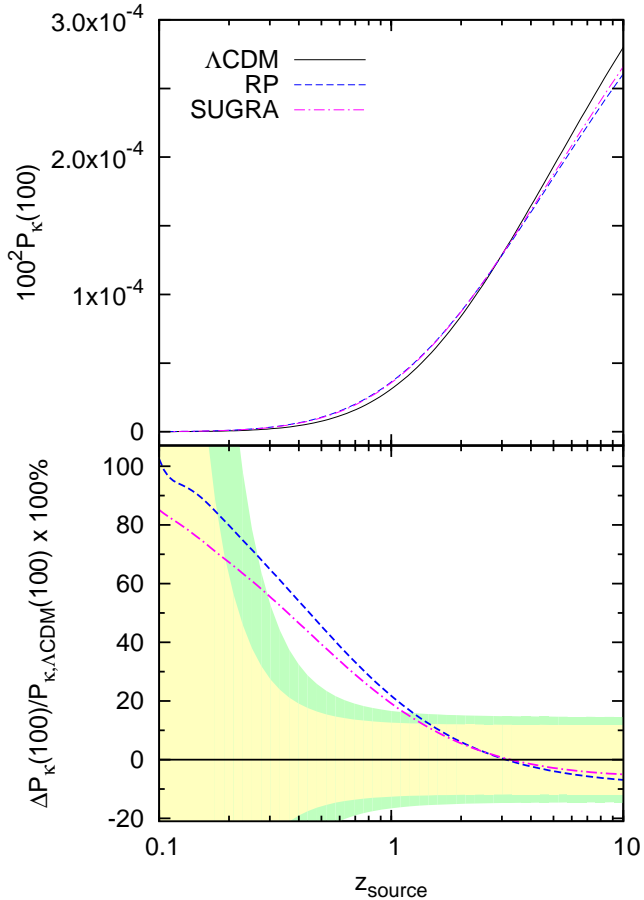, width= 12cm, angle = -90}
\caption{The convergence power at $\ell=100 (\sim 2^\circ)$ as a function of source redshift $z\sub{source}$. The lower panel shows the fractional difference in $P_\kappa(100)$  expressed as a percentage. The outer (inner) contour corresponds to uncertainty \re{intrin} with galaxy number density =\,250 (500)\,arcmin$^{-2}$, ${\cal A}= 20,000$ $(30,000)$\,$\mbox{deg}^2$}
\label{figsource}
\end{figure}

It is interesting to investigate at what redshift would the best prospects for detection of the dark-energy enhancement be. Figure \ref{figsource} shows the convergence power at $\ell=100$ for sources at varying redshifts $(0.1\lesssim z\lesssim10)$. The upper panel shows a clear upward trend for all three dark energy models. The percentage of deviation from the $\Lambda$CDM model is plotted in the lower panel, which shows an interesting tendency for sources at a \ii{lower} redshift to give a higher deviation (at a fixed multipole). By increasing the source redshift, the deviation diminishes and can even disappear altogether (at $z\sim3$ in this case). Beyond this redshift, the weak-lensing signal at scale $\sim2^\circ$ is in fact suppressed compared to the $\Lambda$CDM case.

At first sight the suppression of weak-lensing signal for high-redshift sources appears to be in conflict with figure \ref{figdistance}, which shows that high-redshift objects are further away in the quintessence scenarios, therefore should enhance the weak-lensing signal due to the fact that there is more intervening energy density. However, recall that in equation \ref{pk}, the various density power spectra  $P(\ell/r,r)$ are smaller at larger $r$. For sources at a sufficiently large redshift, the weight function becomes more spread out, thus taking into account more contributions from $P(\ell/r,r)$ at large $r$.  This explains the decreasing curves seen in figure \ref{figsource}. Upon imposing the error contours for the same specifications as in figure \ref{figerrors}, we see a `sweet spot' appearing at around $z\simeq0.5$ where the enhancement level exceeds the intrinsic error by a wide margin. Although this only holds for a single source redshift, we expect our results to be qualitatively correct given a realistic source distribution.

%%%%%%%%%%%%%%%%%%%%%%%%%%
%%%%%%%%%%%%%%%%%%%%%%%%%%

%%%%%%%%%%%%%%%%%%%%%%%%%%
%%%%%%%%%%%%%%%%%%%%%%%%%%
%%%%%%%%%%%%%%%%%%%%%%%%%%

In summary, it is possible to distinguish simple models of dynamical dark energy from the cosmological constant via the boost in the weak-lensing power spectra across linear scales. However, the error contours suggest that it would take a network of at least 30-m class telescopes dedicated to cosmic-shear measurement across a large fraction of the sky to detect the dark-energy enhancement, which can be easily obscured at both high and low $\ell$ by cosmic variance and shot noise. More realistically, there will be additional errors from photometry, systematics in the measurement of galaxy ellipticities, and uncertainties associated with other cosmological parameters. Nevertheless, we expect that they will be subdominant in comparison with the errors calculated in this section. 

\section{Conclusions and discussions}

In this work, we have investigated whether there are observable signatures of dynamical dark energy in weak-lensing measurements. Our basic assumptions are that the background cosmology is a flat FRW universe, containing radiation, dark matter and the simplest type of scalar-field dark energy (with no hot dark matter, baryons or tensor modes). Our technique involves calculating the linear density power spectra by integrating six coupled differential equations, expressed in terms of gauge-invariant density and velocity variables for dark matter, radiation and dark energy (equations \ref{d1}-\ref{v3}). We assume adiabatic initial conditions for all three types of perturbations, and normalise the matter power spectrum (with primordial form $\sim k^{n_s}$) on CMB scales. We base our analysis specifically on two simple models of quintessence, namely the Ratra-Peebles and SUGRA-type potentials, with model parameters that best fit the recent CMB and combined supernova data.

We find that the best-fit quintessence models yield matter power spectra that are hardly distinguishable from that of the $\Lambda$CDM model (figure \ref{figpm}) except on large scales where dark-energy clustering is strongest, as shown in figure \ref{figde}. We believe this is the first time that the dark-energy power spectra for these oft-cited models are explicitly calculated.

When the weak-lensing signals are extracted from the 3D spectra, we found an imprint in the form of an all-scale enhancement in the convergence (or the cosmic shear) power spectrum compared with $\Lambda$CDM cosmology. For sources at redshift 0.5, for instance, the enhancement may be as large as $40\%$ at about $2^\circ$ scale (or 20\% at $z=1$). These enhancements can be attributed to two main reasons, namely i) the change in distance-redshift relation (figure \ref{figdistance}), and ii) the lensing by large-scale dark-energy clusters (figure \ref{figignore}). 

Assuming that the primary sources of errors in the weak-lensing measurements will be from cosmic variance and intrinsic ellipticity dispersion in the shapes of distant galaxies, there are good  prospects for detecting this enhancement for sources at redshift $z\sim0.5$ using future facilities with surveys covering most of the sky, and capable of measuring ellipticities for hundreds of galaxies per arcmin$^{2}$ (figure \ref{figerrors}). On the timescale of a decade, one class of telescope capable of such a survey is the E-ELT, 42m in diameter and including adaptive optics, for which operations are planned to start in 2018. To put this requirement in context, current ground-based surveys typically measure ellipticities for a few tens of galaxies per arcmin$^{2}$ over order 100 square degrees (e.g. CFHTLS-Wide\footnote{http://terapix.iap.fr}), and the highest specification near-future survey DES\footnote{http://www.darkenergysurvey.org} will measure a comparable number density of galaxies over 5000 square degrees.
Here we restricted our consideration to distant galaxies at the same redshift, but for a real survey with sources distributed in redshift, photometric redshift estimates can be obtained using observations in multiple filters. Tomographic information using the auto- and cross-correlations between the cosmic shear in various redshift bins, will give additional leverage in distinguishing between models of dark energy. 

\bbb

\no{\bf{ACKNOWLEDGMENTS}}

\mmm
\noindent
SC is supported by Lincoln College, Oxford. LJK is supported by the Royal Society. We are grateful to Antony Lewis for reading the manuscript and for providing many helpful suggestions.

\bibliographystyle{mn2e}
\bibliography{lensing}

\end{document}